# Magneto-optical imaging of thin magnetic films using spins in diamond

David A. Simpson[1,2,*], Jean-Philippe Tetienne[1,3], Julia McCoey[1], Kumaravelu Ganesan[1], Liam T. Hall[1], Steven Petrou[2,4,5], Robert E. Scholten[1], Lloyd C. L. Hollenberg[1,2,3]

[1]School of Physics, University of Melbourne, Parkville, 3052, Australia
[2]Centre for Neural Engineering, University of Melbourne, Parkville, 3052, Australia
[3]Centre for Quantum Computation and Communication Technology, University of Melbourne, Parkville, 3052, Australia
[4]Florey Neuroscience Institute, University of Melbourne, Parkville, 3052, Australia
[5]Centre for Integrated Brain Function, University of Melbourne, Parkville, Victoria, Australia
*corresponding author simd@unimelb.edu.au

Imaging the fields of magnetic materials provides crucial insight into the physical and chemical processes surrounding magnetism, and has been a key ingredient in the spectacular development of magnetic data storage. Existing approaches using the magneto-optic Kerr effect (MOKE), x-ray and electron microscopy have limitations that constrain further development, and there is increasing demand for imaging and characterisation of magnetic phenomena in real time with high spatial resolution. In this work, we show how the magneto-optical response of an array of negatively-charged nitrogen-vacancy spins in diamond can be used to image and map the sub-micron stray magnetic field patterns from thin ferromagnetic films. Using optically detected magnetic resonance, we demonstrate wide-field magnetic imaging over 100×100 µm$^2$ with a diffraction-limited spatial resolution of 440 nm at video frame rates, under ambient conditions. We demonstrate a novel all-optical spin relaxation contrast imaging approach which can image magnetic structures in the absence of an applied microwave field. Straightforward extensions promise imaging with sub-µT sensitivity and sub-optical spatial and millisecond temporal resolution. This work establishes practical diamond-based wide-field microscopy for rapid high-sensitivity characterisation and imaging of magnetic samples, with the capability for investigating magnetic phenomena such as domain wall and skyrmion dynamics and the spin Hall effect in metals.

Magnetic imaging techniques [1] are generally characterized by their spatial and temporal resolution, but criteria such as sensitivity, field disturbance, sample damage, field of view, cost, and ease of use, are critical for broad applicability which ultimately drives future development of our understanding of magnetism in advanced materials and applications. Electron and x-ray microscopy can provide high spatial resolution down to a few nanometres [2-4], but are time-consuming, and require expensive complex apparatus, careful sample preparation, and a high vacuum environment. Magnetic force microscopy (MFM) [5, 6] and magneto-optical Kerr microscopy (MOKE) [7, 8] allow rapid characterisation of magnetic devices, but MFM is inherently slow and is not suited to imaging fragile magnetisation states due to its invasive magnetic tip. MOKE, on the other hand, is a non-invasive optical technique which has been used to great effect in furthering our understanding of the spin Hall effect [9] and more recently the formation of magnetic skyrmion bubbles under ambient conditions [8], but is limited to samples that exhibit a strong Kerr response.

Solid state spin systems offer a new approach for magnetic imaging, based on the sensitivity of quantum spin states to external magnetic fields. In particular, the negatively charged nitrogen-vacancy (NV) centre in diamond has demonstrated competitive magnetic sensitivity in ambient room temperature environments [10]. Diamond-based imaging techniques have been developed for biological cells [10, 11], current-carrying wires [12-14] and paramagnetic molecules [15]. Here we introduce a new approach for imaging solid state magnetic devices, overcoming limitations in current state-of-the-art magnetic imaging techniques.

More specifically, we use the magnetic sensitivity of NV centres in diamond to image the stray magnetic fields in thin ferromagnetic films, with commercial recording media as an example. We demonstrate temporal and spatial resolution of 10 ms and 440 nm respectively. Our technique is applicable to any magnetic material with a stray magnetic field and can be operated with high throughput by simply placing the material in contact with the diamond imaging chip under normal ambient conditions. The apparatus consists of a conventional commercial wide-field fluorescence microscope with a diamond imaging chip onto which the magnetic sample is mounted. The magnetic contrast is obtained from the fluorescence signal emitted by NV defect centres engineered beneath the diamond surface. We present three distinct imaging modalities which can be used independently or in combination. In the first, we use the optically detected magnetic resonance (ODMR) of the NV spins to resolve the iso-magnetic field lines from the magnetic bits in the absence of an applied magnetic field. In the second example, we describe an imaging protocol with an applied magnetic field to quantitatively map the stray magnetic field from each individual bit over the entire field of view and compare the results to theoretical simulations of the recording media. Finally, we demonstrate a new all-optical imaging modality based on NV spin relaxation contrast, particularly useful for imaging strong off-axis magnetic fields.

All three imaging modalities were performed on the same wide-field magnetic imaging microscope, see Fig. 1a. The magnetic sample of interest requires no special treatment and is simply placed in contact with a 2x2 mm$^2$ diamond imaging chip. A two-dimensional array of near-surface NV centres was engineered in the diamond by nitrogen ion implantation and subsequent annealing. The implantation energy of 20 keV resulted in a mean NV depth of approximately 30 nm. The NV array is illuminated by a 532 nm green laser and the resulting red fluorescence (650-750 nm) is imaged onto an sCMOS camera, see Fig. 1b. Using a Nikon ×40, 1.2 NA oil objective lens, we obtain a field of view of 100×100 μm$^2$, and operate with an optical power density of 30 W/mm$^2$. Microwave (MW) excitation is provided by a resonator lithographically patterned onto a glass coverslip placed below the diamond imaging chip for acquisition of the ODMR spectrum from the NV spins in the array.

Figure 1c shows a typical ODMR spectrum obtained from the integrated signal over the entire field of view with and without an external magnetic field. The ground state electron spin sublevels $m_s = \pm 1$ of each NV centre are Zeeman split in the presence of a local magnetic field, resulting in a frequency shift in the spin levels of $\Delta f = \pm \gamma_e B_{NV}/2\pi$ where $\gamma_e$ is the electron gyromagnetic ratio and $B_{NV}$ is the magnetic field projection along the NV symmetry axis [11]. The NV centres are randomly oriented along the four <111> crystallographic axes of diamond with average separation of 20 nm, assuming 1% conversion efficiency of [N] to [NV]. The ODMR spectrum therefore exhibits four pairs of resonance lines corresponding to the $B_{NV,i=1..4}$ magnetic field projections. The ODMR sensitivity to local magnetic fields provides a contrast mechanism which we use to provide imaging of the stray magnetic field associated with the sample, in this case commercial recording media.

## Results

### 1. Wide-field imaging of iso-magnetic field lines in thin ferromagnetic films

A thin ferromagnetic film sample was obtained by cutting a segment of recording media from a commercial magnetic hard disk drive. The magnetic film was encoded with random magnetic data on bits 14 x 1.5 µm determined independently by MFM (see *Supplementary Information*). The magnetic bits are magnetised in-plane, parallel to the tracks, with random sign. The magnetic sample (5x5 mm$^2$) was simply placed face down on the diamond imaging chip. Due to surface imperfections a distance between the magnetic film and the NV array of 1 µm was observed over the imaging region. Since this stand-off distance is of the order of the resolution of the microscope, it has negligible impact on the spatial resolution but reduces the measured stray magnetic field strength. For improved distance control, the magnetic device could be directly fabricated on the diamond surface, but this would likely not be required in many applications, thus allowing high throughput and characterisation of a range of magnetic materials.

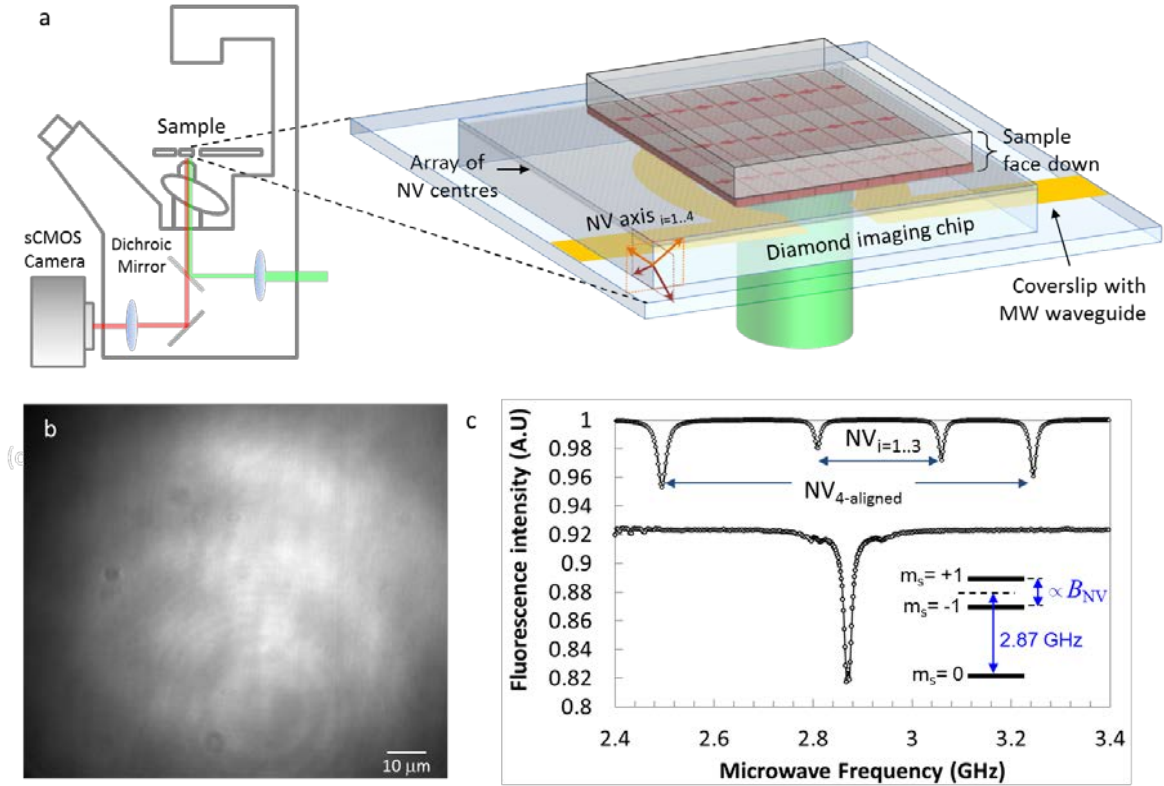

Figure 1: Experimental arrangement for diamond-based wide-field magnetic imaging. (a) Schematic diagram of the instrument, illustrating the inverted microscope with green laser excitation and an sCMOS camera to image the fluorescence from a two-dimensional array of NV centres implanted in diamond. The exploded assembly at right illustrates the diamond imaging chip mounted on a glass coverslip equipped with a microwave (MW) resonator. The magnetic sample is placed face down onto the diamond. (b) Raw fluorescence image from the NV array. (c) ODMR spectra of the NV array integrated over the entire field of view with (upper) and without (lower) applied magnetic field. The applied field (13.4 mT) was aligned along one particular NV axis with the other three NV axes experiencing the same magnetic field projection. This results in a pair of ODMR lines for the aligned NV centres (NV$_{aligned}$) and a second set of ODMR lines for the misaligned NV centres.

To image the iso-magnetic field lines, a single microwave frequency $f$ is applied to the imaging array which is resonant with a spin transition of the NV centre only when the local magnetic field $B_{NV}$ satisfies the condition: $\Delta f = f - f_D = \pm \gamma_e B_{NV}/2\pi$, where $f_D$ is the resonance frequency in zero field. At resonance, the microwave field induces a decrease of up to 10% in the NV fluorescence intensity (Fig. 1c). The recorded fluorescence image then reveals iso-magnetic field contours corresponding to a given value of $B_{NV}$ as lines of low intensity. To reduce the effect of illumination inhomogeneity (see Fig. 1b), we divide images with and without MW excitation. A typical iso-magnetic field image is shown in Fig. 2a, where a MW frequency of 2.79 GHz was applied, corresponding to a field projection of $B_{NV} = \pm 2.9$ mT. Since each pixel contains the signal from a large number of NV centres located in four distinct orientations, see Fig. 1a, the method produces magnetic field contrast whenever the stray magnetic field projection is resonant with the probe MW frequency along each NV axis.

The individual bits are clearly observed over the entire field of view. The acquisition time for the images shown in Fig. 2a and 2c was 48 seconds. Integration over this time period reduces the shot noise in the fluorescence measurement, resulting in clearly resolved iso-magnetic field lines. A typical line cut through a single iso-magnetic field line is shown in Fig. 2b demonstrating a magnetic imaging resolution of 440 ± 20 nm, consistent with the diffraction limit of our wide-field 1.2 NA microscope operating at λ = 650 – 750 nm: $(1.22\ \lambda/(2NA) = 380\text{nm}$ at 750nm.

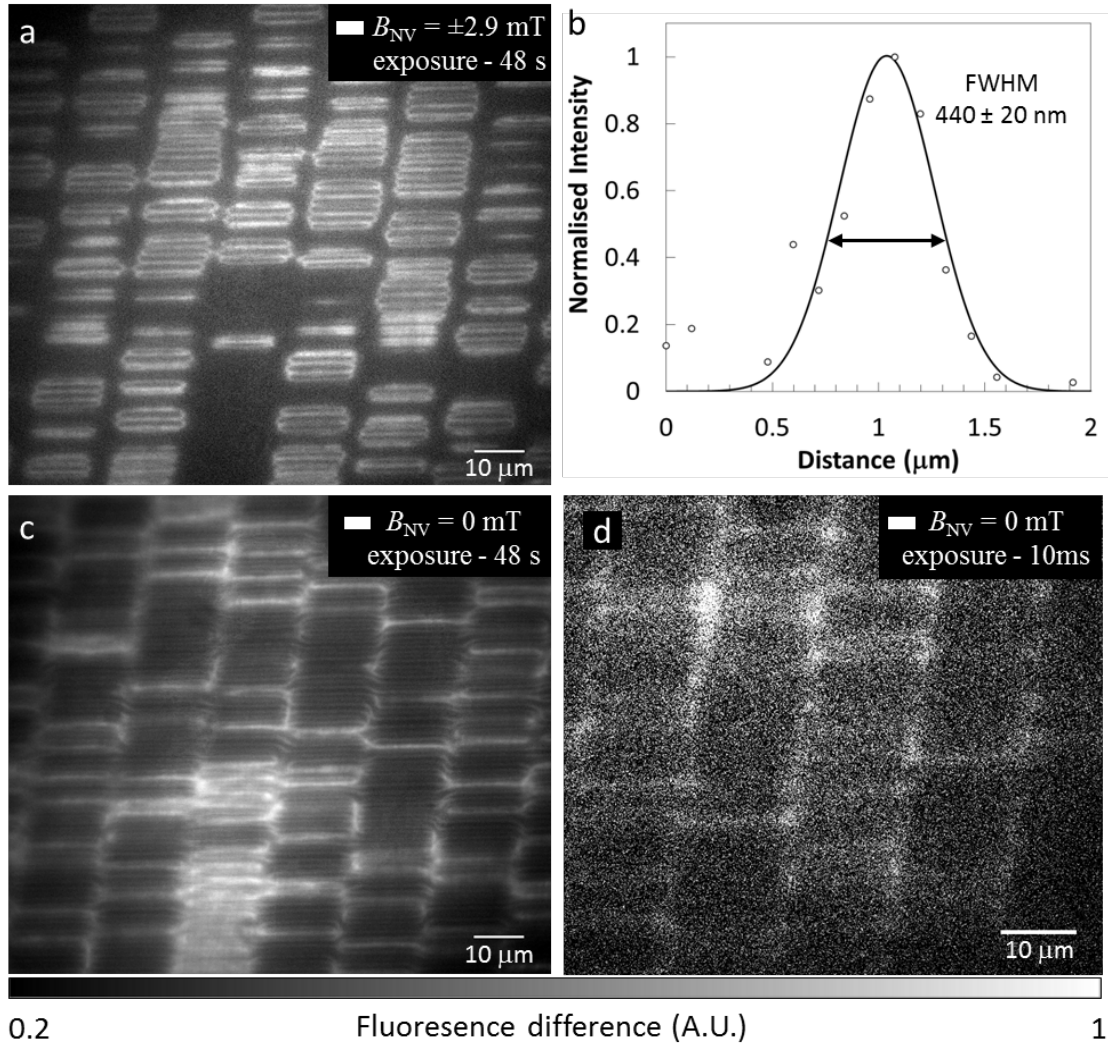

Figure 2: Wide-field imaging of the iso-magnetic field contours of an array of magnetically encoded bits. (a) Iso-magnetic field image showing in white the contours of the stray field projection $B_{NV,i=1..4}$ = ±2.9 mT (probe MW frequency 2.79 GHz). The image is formed by dividing two fluorescence images with and without MW excitation. The imaging area is 100×100 μm² and the integration time is 48 s. (c,d) Iso-magnetic field image showing in white the contours of the projection $B_{NV,i=1..4}$ = 0 mT (probe MW frequency 2.87 GHz). The integration time was 48 s and 10 ms for (c) and (d), respectively. The two images correspond to separate areas of the sample.

To probe the temporal response of the system, the integration time per image was reduced until the magnetic bits could no longer be resolved. For this experiment a MW frequency of 2.87 GHz corresponding to a projection $B_{NV}$ = 0 was applied, so that all NV centres contribute to the signal in the regions of zero magnetic field, thereby maximizing the contrast of the resulting image. Figure 2d shows a magnetic image of the iso-magnetic field lines acquired in 10 ms. This temporal resolution is several orders of magnitude faster than MFM, and comparable to that of wide-field MOKE microscopy [8].

Another important characteristic of the microscope is its sensitivity, which limits the smallest number of Bohr magnetons in the sample that can be detected for a given integration time. The magnetic field sensitivity of the imaging chip is dictated by the width and contrast of the Lorentzian ODMR peaks, with the optimal sensitivity given by [12]

$$\eta_{dc} = \frac{4\,h\delta}{3\sqrt{3}\cdot g_{NV}\gamma_{e_B} R\sqrt{n}}$$

where $\delta = 6$ MHz is the full width at half maximum of the ODMR peak, and the ESR contrast $R = 0.1$. The number of photons per detection volume (4x4 pixels, 275 × 275 nm) is $n = 1.2 \times 10^6 \text{s}^{-1}$ giving a magnetic sensitivity of 1.5 $\mu$T/$\sqrt{\text{Hz}}$ per detection volume, consistent with the DC sensitivity reported elsewhere in similar ensemble samples [13-15]. It should be noted that with modest diamond material improvements the projected field sensitivity can be significantly enhanced. For example, in instances where the stray magnetic field decays over length scales of microns, much deeper and more uniform NVs can be engineered with much higher [N] to [NV] conversion efficiencies > 10% [14, 16]. In this case, the $\sqrt{n}$ improvement along with the reduction in the ODMR line width can result in projected DC sensitivity of order nT/$\sqrt{\text{Hz}}$ [17]. In most solid state magnetic device imaging applications the magnetic fields are of order tens of $\mu$T, therefore we expect resolution of sub-micron magnetic structures on millisecond timescales.

**2. Quantitative stray magnetic field mapping.**

To quantify the stray field from the array of magnetic bits, a matrix of ODMR spectra was acquired by recording fluorescence images while sweeping the MW frequency. The ODMR spectrum exhibits in general four pairs of resonance lines with Zeeman splitting corresponding to the field projection along each of the four NV axes. In principle, the vector magnetic field can be reconstructed from the ODMR matrix [13]. The procedure is time consuming and challenging for small (< mT) magnetic fields when the four line pairs cannot be readily resolved. An alternate approach is to apply an external magnetic field along one particular NV axis to separate the four orientations (Fig. 1d), allowing measurement of the stray field projected along a particular NV axis ($B_{NV}$). Both the sign and magnitude of $B_{NV}$ with respect to the background field can be obtained without assumptions regarding the source of the magnetic field.

Figure 3a shows a quantitative magnetic field map of the magnetic bits. In the top part of the image, the measured field periodically oscillates between +0.5 and -0.5 mT along the data tracks, with a period of 1.47 μm (Fig. 3b). This period closely matches the bit spacing obtained from MFM (1.44 μm), indicating that in this region of the sample the bits are magnetised according to a regular sequence 010101. The white regions of the image highlight the zero magnetic field regions between each bit and along the data tracks. Note that the full vector stray field from the magnetic sample could be formed by repeating the measurement with the external magnetic field aligned along each of the other three NV axes [18].

To assess the imaging performance of our microscope we compare the experimental data to a micromagnetic model of the device. Our model calculates the stray magnetic field from an array of uniformly magnetised rectangular elements [19] which is then projected along the NV axis used in the experiment (see Methods). The only unknown parameter in the simulation is the stand-off distance of the NV array with respect to the magnetic sample. The simulation best reproduces the amplitude of the field in the experimental data (0.5 mT) for a stand-off distance of 1 μm. The resulting image (Fig. 3c) and detailed field profile (Fig. 3b) are in excellent agreement with the experimental data.

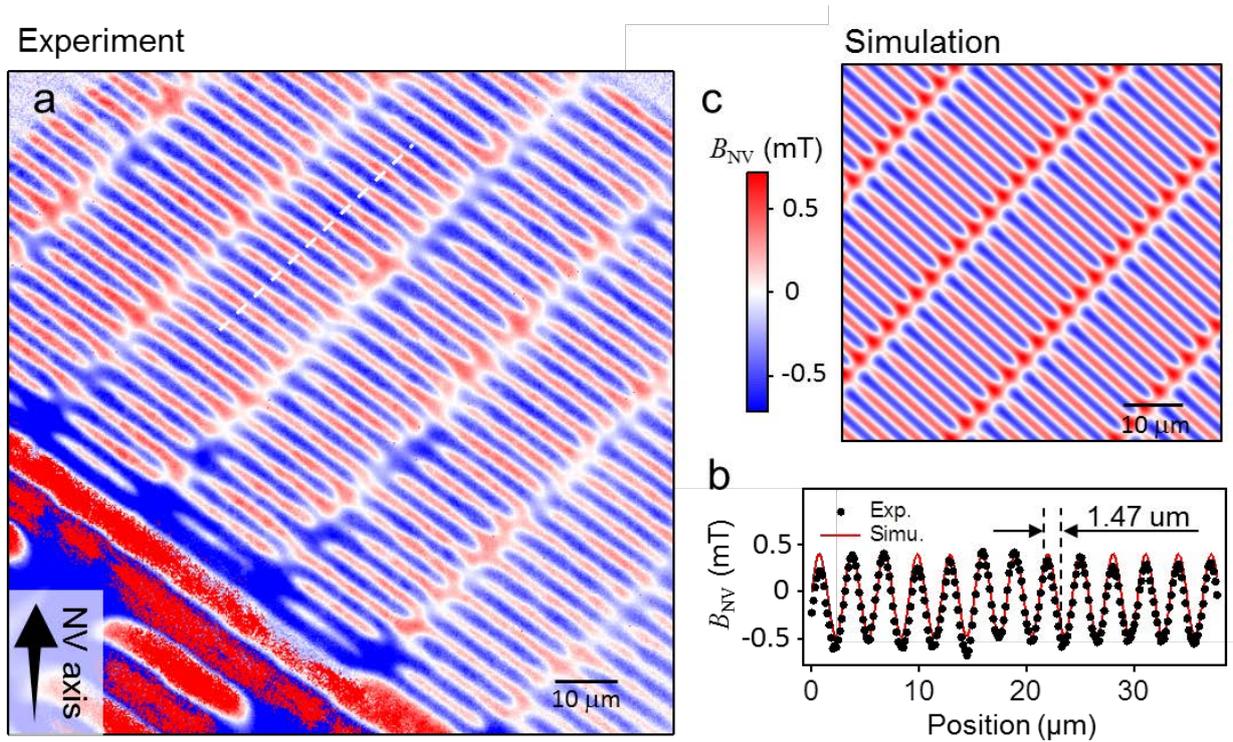

Figure 3: Wide-field quantitative magnetic imaging. (a) Measured map of the field projection $B_{NV}$ along a particular NV axis, as depicted by the black arrow. A background field of 13.4 mT aligned with the NV axis was applied during the acquisition. The total acquisition time is 24 min. (b) Line cut taken along the dashed line in (a), showing that the magnetic field oscillates with a period of 1.47 μm. (c) Simulated image corresponding to an infinite array of bits with a regular magnetisation sequence 010101. A stand-off distance of 1 μm gives good quantitative agreement with (a). A line cut taken along a track in (c) is also shown in (b) for direct comparison with the experiment.

## 3. All-optical spin relaxation contrast imaging.

Finally, we demonstrate a new all-optical magnetic imaging technique based on the NV spin relaxation contrast. In the presence of off-axis magnetic fields (> 5 mT), the efficacy of the NV spin polarisation in the imaging modalities described above is significantly reduced because the natural quantisation axis for the NV centre is then defined by the background magnetic field rather than by the intrinsic crystal field, and the spin sublevels $m_s = 0$ and $\pm 1$ are no longer eigenstates of the unperturbed NV system. In this case, mixing of the spin state populations causes a reduction in the spin polarisation efficiency, and thereby the fluorescence intensity under optical pumping [20]. This spin mixing hampers ODMR-based imaging protocols at high magnetic field strengths, but also provides an alternate approach to magnetic imaging via the optical response of the NV centre.

Previous all-optical nanoscale magnetic imaging has been achieved with single NV centres in nanodiamonds attached to scanning atomic force microscope tips, with magnetic contrast being derived from the change in fluorescence from a single NV centre due to spin mixing in the excited states [19]. When scaling this technology to NV ensembles for wide-field imaging, it becomes increasingly difficult to characterise small variations in the fluorescence, since the NV distribution and pump excitation are not uniform over the large field of view. To overcome this issue, we developed a new imaging protocol based on the fluorescence contrast of the spin relaxation time ($T_1$) of the imaging array. $T_1$ relaxation imaging has been proposed [21] and recently demonstrated as an approach to image weak fluctuating magnetic fields [22, 23]. However, in the presence of high off-axis magnetic fields, the reduced spin polarisation caused by spin mixing complicates the measurement of the $T_1$ relaxation time. Instead we describe here a technique which takes advantage of the spin mixing effect on $T_1$ and show that the contrast of the $T_1$ relaxation curve can be used to quantify regions of high off-axis magnetic fields. To illustrate the relaxation contrast imaging approach we first implement the traditional spin relaxation time protocol shown in Fig. 4a, where the $T_1$ relaxation curve is mapped at each pixel. Figure 4c shows two typical $T_1$ relaxation curves denoted as $S_{x1y1}(\tau)$ and $S_{x2y2}(\tau)$ corresponding to two distinct pixel regions of high and low off-axis magnetic fields as shown in the ODMR-based magnetic image in Fig. 4d. The effect of spin mixing on the contrast of the spin relaxation curve $S_{xy}(\tau)$ can be seen clearly in Figure 4c with the $S_{xy}(\tau)$ in areas of high off-axis magnetic fields showing a reduced fluorescence contrast. By probing the fluorescence signal at two distinct spin relaxation times, $S_{xy}(\tau_1)$ and $S_{xy}(\tau_2)$ the spin relaxation contrast can be obtained over the entire imaging area by summing the normalised fluorescence signal from each measurement cycle $\sum_{i=0}^{n} S_{xy}(\tau_1)/S_{xy}(\tau_2)$. The spin relaxation contrast sequence is shown in Fig. 4b. The two relaxation time points used in this example were $\tau_1$=10 µs and $\tau_2$=860 µs as shown in Fig. 4c and the magnetic image was formed after integrating for $2 \times 10^6$ cycles, 30 mins. Figure 4d shows the all-optical spin relaxation contrast image of the magnetic recording media, where white contours corresponds to high off-axis magnetic fields. Calculations of the NV spin dynamics show a contrast change of a few % corresponds to off-axis magnetic field strengths $\geq$ 5mT [20]. The relaxation contrast image is in excellent agreement with the corresponding low magnetic field image obtained using the ODMR-based protocol shown in Fig. 4d. The comparison illustrates the complementarity between the two methods: ODMR reveals the low-field regions, while spin relaxation contrast imaging highlights the large off-axis magnetic fields. It should be noted that the magnetic sample used in this study provides modest off-axis magnetic fields of order 5mT providing ~2% relaxation contrast. In cases where the off-axis stray magnetic fields are considerably larger >20mT the spin contrast will increase up to 20% [20] which will allow for much faster acquisition rates of order seconds. This would enable dynamic imaging of high magnetic fields using the optical response of the NV spins.

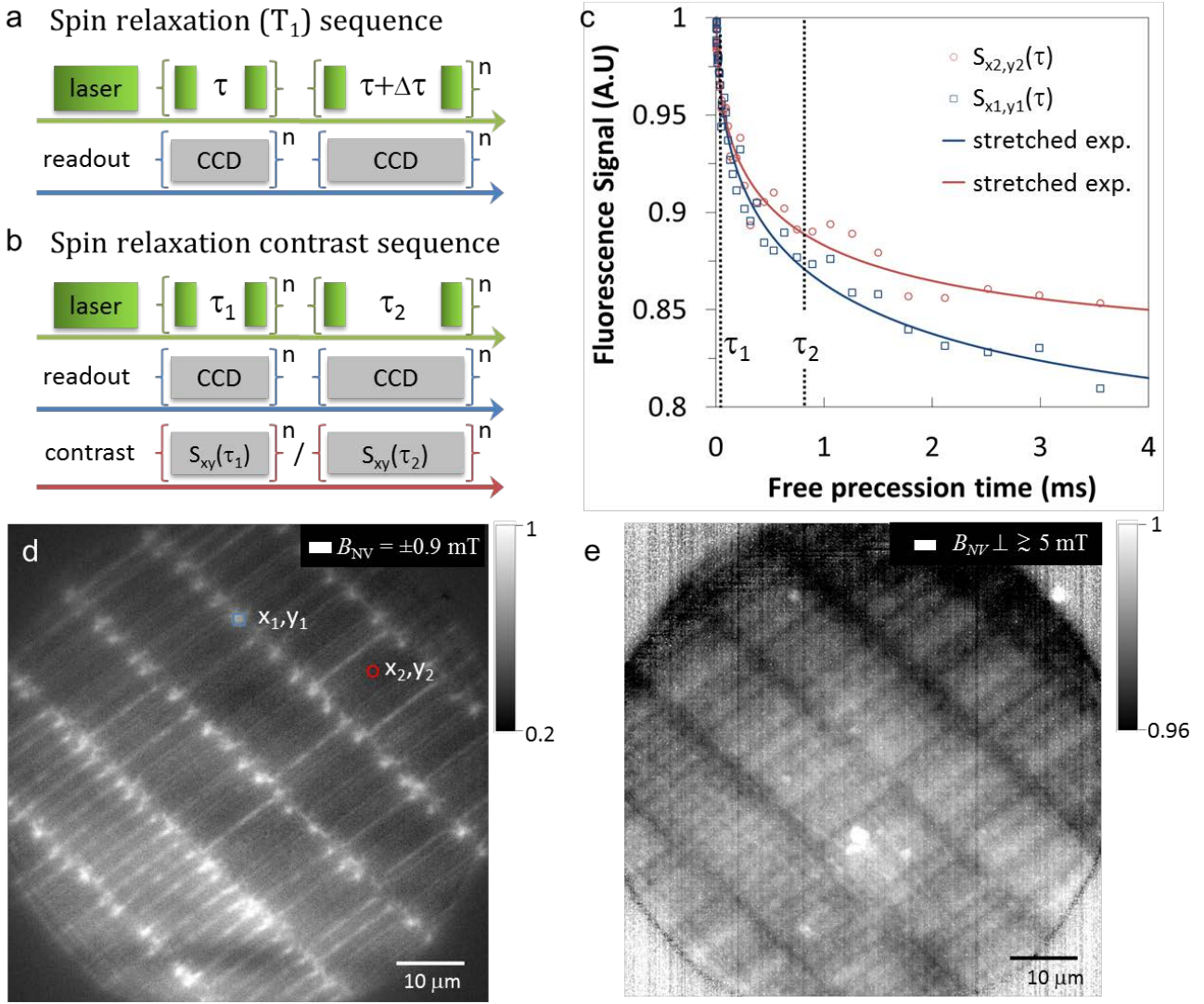

Figure 4: All optical wide-field spin relaxation contrast imaging. (a) Experimental pulse sequence used to measure the spin lattice relaxation time $T_1$. The optical excitation pulses are shown in green with the sCMOS exposure times in grey. (b) All optical spin relaxation contrast imaging protocol. Magnetic contrast is obtained by normalising the fluorescence signal $S_{xy}(\tau_1)$ and $S_{xy}(\tau_2)$. (c) $T_1$ relaxation curves from two distinct 4x4 pixels areas shown in (d), revealing a difference in spin relaxation contrast for areas of high and low off-axis magnetic fields. (d) Iso-magnetic field image obtained from the magnetic recording media. The white contours indicate regions of low magnetic field projection $B_{NV} = \pm 0.9$ mT (e) All optical spin relaxation contrast image of the same region as (d). The acquisition time is 30 minutes with the white contours showing regions of high off axis magnetic fields $B_{NV}\perp \gtrsim 5$ mT.

## Discussion

In this work we have shown that NV defects in diamond provide the basis of a sensitive optical wide-field magnetic microscope for the characterisation of magnetic materials with high spatial resolution and video frame rates. The magnetic sensitivity to DC fields is 1.5 µT/$\sqrt{Hz}$ and is capable of imaging the stray magnetic field from thin ferromagnetic films in 10 ms under ambient conditions. The method would allow magnetic imaging of a single 10 nm magnetic skyrmion in an ultrathin ferromagnet, at a stand-off distance of 300 nm or less [24], with 1 second acquisition time. Larger skyrmions (up to 1 µm in diameter) [8] could be imaged within the 10 ms temporal resolution limit of our existing instrument.

The diffraction limited spatial resolution is achieved over a large field of view (100×100 µm$^2$), in the absence of any external magnetic field which may impact the magnetisation of the sample under investigation. The spin relaxation contrast from NV centres in diamond also enables a new all-optical imaging approach suitable for magnetic structures that exhibit large off-axis magnetic fields, for example super-conducting vortices[25], again over wide fields of view. Since our imaging magnetometer relies on mapping the stray magnetic field rather than on a direct interaction with the sample magnetisation, it is applicable to all samples regardless of their shape (thickness, surface roughness) and composition. This is in contrast to existing techniques which impose constraints on the type of sample that can be imaged, particularly for non-magnetic metals, where direct imaging of spin injection or the spin Hall effect have not been possible [26].

The minimal requirements on sample and instrument preparation will allow high throughput characterisation of magnetic samples or devices. A number of improvements can be anticipated. Both the temporal resolution and sensitivity can be improved by increasing the fluorescence intensity per detection volume, by increasing the density of NV centres and by optimising the photon collection efficiency. With an [N] to [NV] conversion efficiency improvement from 1 to 10% [27], millisecond temporal resolution and magnetic sensitivity of sub-µT will be possible. Improved spatial resolution may also be possible by implementing super-resolution schemes such as stimulated emission depletion [28] or ground state depletion microscopy [29], although at the expense of longer acquisition times. There are many important applications, for example dynamic magnetic phenomena such as the motion of magnetic domain walls and skyrmions under current, magnetic field or thermal fluctuations [8, 30, 31] and exploration of new materials, particularly metallic and non-magnetic materials, where existing magnetic imaging techniques are lacking. Diamond-based wide-field microscopy thus provides a new pathway to study these technologically important magnetic phenomena.

## Methods

### Materials

The imaging sensor used in this work is engineered from electronic grade Type IIa diamond (Element 6). The diamonds were thinned and repolished to a 4 x 4 x 0.1 mm crystal (DDK, USA) and then laser-cut into 2x2 mm sensors. The single crystal diamond was then implanted with $^{15}$N atoms at 20 keV to a dose of 1x10$^{13}$ ions/cm$^2$. The implanted sample was annealed at 1000 °C for three hours and acid treated to remove any unwanted surface contamination. The density of NV centres post annealing was 1x10$^{11}$ NV/cm$^2$, estimated by comparing the intensity from a single NV centre in diamond with that obtained from the NV ensemble.

### Imaging

The wide-field imaging was performed on a modified Nikon inverted microscope (Ti-U). Optical excitation from a 532 nm Verdi laser was focused ($f$ = 300mm) onto an acousto-optic modulator (Crystal Technologies Model 3520-220) and then expanded and collimated (Thorlabs beam expander GBE05-A) to a beam diameter of 10 mm. The collimated beam was focused using a wide-field lens ($f$ = 300mm) to the back aperture of the Nikon x40 (1.2 NA) oil immersion objective via a Semrock dichroic mirror (Di02-R561-25x36). The NV fluorescence was filtered using two bandpass filters before being imaged using a tube lens ($f$ = 300mm) onto a sCMOS camera (Neo, Andor). Microwave excitation was provided by an Agilent microwave generator (N5182A) and switched using a Miniciruits RF switch (ZASWA-2-50DR+). The microwaves were amplified (Amplifier Research 20S1G4) before being sent to the microwave resonator. A Spincore Pulseblaster (ESR-PRO 500MHz) was used to control the timing sequences of the excitation laser, microwaves and sCMOS camera and the images where obtained and analysed using custom LabVIEW code. The excitation power density used for imaging was 30 W/mm$^2$ and all images were taken in an ambient environment at room temperature. The magnetic sample used for imaging was cut from a commercial Western Digital 1 GB magnetic hard drive.

**Stray magnetic field simulations**

The sample is modelled as an array of bits where each bit is a parallelepiped with uniform magnetisation. The size of the bits is 14 μm x 1.5 μm x 15 nm, and the pitch of the data tracks is 17 μm, which means that there is a gap of 3 μm between bits of adjacent tracks. These parameters were taken directly from the MFM images. The magnetisation of each bit is assumed to lie in the plane of the magnetic film, mainly parallel to the direction of the data tracks. A tilt of 25° with respect to this main direction is added to account for the effect of the applied background magnetic field. The saturation magnetisation is taken to be 0.5 x 10$^6$ A/m. Using analytical formulae of the field generated by one bit [19], one can sum over the array to obtain the vector field map at a given distance d from the sample. The field is then projected onto the NV axis used in the experiment, which gives the map of $B_{NV}$. Finally, a convolution with a Gaussian lineshape of 440 nm full width at half maximum is applied to account for the diffraction limited optical response of the microscope. The stand-off distance $d$ was varied in order to find the best match with the experiment.